\documentclass[12pts]{elsart}
\usepackage{graphicx}
\usepackage{amssymb}  
\usepackage{color} 
\begin{document}
\begin{frontmatter}
\title{Detailed Studies of Pixelated CZT Detectors Grown with the Modified Horizontal Bridgman Method}
\author[a1]{I. Jung \corauthref{cor1}}
\ead{jung@physics.wustl.edu}
\corauth[cor1]{I. Jung}
\author[a1]{H.~Krawczynski}
\author[a2]{A.~Burger}
\author[a2]{M.~Guo}
\author[a2]{M.~Groza}
 \address[a1]{Washington University 
   Department of Physics
   1 Brookings Dr., CB 1105
   St Louis
   MO 63130}
\address[a2]{Fisk University 
  Department of Physics         
  1000 Seventeenth Avenue North 
  Nashville, TN 37208-3051     }
\begin{abstract}
The detector material Cadmium Zinc Telluride (CZT) achieves excellent 
spatial resolution and good energy resolution over a  broad energy range, 
 several keV up to some MeV.  
  Presently, there 
  are two main methods to grow CZT crystals, the Modified High-Pressure 
  Bridgman (MHB) and the High-Pressure Bridgman (HPB) process. 
  The study presented in this paper is based on MHB CZT substrates from 
  the company Orbotech Medical Solutions Ltd. \cite{Orbotech}.
   Former studies have shown that 
  high-work-function materials on the cathode side reduce the 
  leakage current and therefore improve the energy resolution at lower 
  energies. None of the studies have emphasized on the anode contact material.
Therefore, we present in this paper the result of a detailed
  study in which for the first time the cathode material was kept constant and the 
anode material was varied. We used four different anode materials : 
  Indium, Titanium, Chromium and Gold, metals with work-functions between
  4.1~eV and 5.1~eV. 
The detector size was  2.0$\times$2.0$\times$0.5~cm$^3$ 
with 8$\times$8 pixels and a pitch of 2.46~mm.
The best performance was achieved with the low work-function materials Indium and Titanium with energy 
resolutions of 2.0~keV (at 59~keV) and 1.9~keV (at 122~keV) for Titanium  and 2.1~keV (at 59~keV) and 2.9~keV (at 122~keV)
 for Indium.  Taking into account the large pixel pitch of 2.46~mm, 
these resolutions are very competitive in comparison to those achieved with 
detectors made of material produced with the more expensive conventional HPB method. We present a detailed comparison of 
our detector response with 3-D simulations. The latter comparisons allow us to determine 
the  mobility-lifetime-products ($\mu\tau$-products) for electrons and holes. 
Finally, we evaluated the temperature dependency of the detector
performance and $\mu\tau$-products.
For many applications temperature 
dependence is important, therefore, we extended the scope of our study to temperatures as low as -30$^{\circ}$C. There are 
two important results. The breakdown voltage increases with decreasing temperature, and electron mobility-life-time-product decreases by about 30\% over a range from 20$^{\circ}$C to -30$^{\circ}$C. The latter effect causes the energy resolution to deteriorate, but  the concomitantly increasing breakdown voltage makes it possible to increase the applied bias voltage and restore  the full performance.
\end{abstract}
\end{frontmatter}
\section{Introduction}
\label{Introduction}
Cadmium Zinc Telluride (CZT) has emerged as the material of choice for the detection of hard X-rays and soft gamma-rays with excellent position and energy resolution and without the need for cryogenic cooling. The high density of CZT ($\rho\,\simeq$~5.76 g/cm$^3$) and high average atomic number ($\simeq$~50) result in high stopping power and in a large cross section for photoelectric interactions.

The main application of CZT detectors is the detection of photons in the 10~keV to $\sim$1~MeV energy range. CZT has a major impact in various fields including medical imaging, homeland security applications, and space-borne X-ray and gamma-ray astronomy.
To give a few examples of the latter, the Swift mission launched in 2004
carries a wide field of view X-ray telescope for the discovery of gamma-ray 
bursts (GRBs) in the energy range from 15~keV to 150~keV \cite{swift}.
This Burst Alert Telescope makes use of the coded mask technique to localize 
GRBs with an accuracy of 1-2~arcmin. It is built out of an array of 32,768 
co-planar 2$\times$4$\times$4~mm$^3$ CZT detectors covering a total area of 
0.5~m$^2$. The proposed EXIST (Energetic X-ray Imaging Survey Telescope) 
mission \cite{Grindlay} also uses the coded mask approach. 
With 15,000 pixelated CZT detectors, each with a volume of 2.0$\times$2.0$\times$0.5~cm$^3$ and with 16$\times$16 pixels, an angular resolution of 5 arcmin will be achieved and the energy range between 10~keV and 600~keV will be covered.\\
In this paper, we present a detailed study of CZT grown with the Modified 
Horizontal Bridgman (MHB) process by the company Orbotech Medical Solutions 
Ltd. \cite{Orbotech} (previously Imarad). This growth technique gives uniform 
substrates at high yields and thus modest costs. One of the disadvantages of 
the MHB technique is a somewhat lower bulk resistivity of $10^9\,\Omega\,$cm 
compared to $10^{10}\,\Omega\,$cm obtained with the more conventional High 
Pressure Bridgman (HPB) process. This results in higher leakage currents and therefore degrades the energy resolution at lower energies ($<$ 100~keV).

Several authors recognized that it is possible to reduce dark currents in 
MHB detectors by using {\sl p}-type-intrinsic-{\sl n}-type (PIN) and 
metal-semiconductor-metal (MSM) 
contacts \cite{Nemirovski:01,Vadawale:04,Nari:98,Nari:00,Nari:02}. 
Both schemes, PIN and MSM, can indeed reduce the dark currents by factors 
$>$10 and improve on the energy resolution of the detectors at low 
($<$ 100 keV) energies. Encouraged by the result of these authors, we 
experimented with different cathode and anode materials on several MHB 
substrates, optimizing for the first time the cathode and anode contacts 
separately. We discussed a comparison of different  cathode materials in an earlier paper \cite{Jung:05}.  Gold (Au) was among the metals
yielding the best performance in terms of dark current suppression and energy
resolution as cathode contact material.
With the objective to finalize our studies on contact materials, we present in this paper 
the results on different anode contact materials with an Au cathode and added a study 
of the performance at lower temperatures which is especially important for space borne applications. We derived $\mu\tau$-products of electrons and holes needed for detector simulations. These reproduce the measured data well, and can be used to further improve detector performance and design.

The paper is structured as follows: \\
In section \ref{sec:detectorfabrication}, origin and some general properties of our substrate 
are presented. We explain how we used this substrate to fabricate detectors suited for our measurements and give a short description of our experimental setup. Then in section \ref{sec:char}, we discuss crystal inhomogeneities and present the results of photoluminescence mapping of the Zn content and of infrared transmission microscopy. Subsequently, in section \ref{sec:contact}, the choice of the contact materials is explained, and the design of the detectors we used for comparison measurements is described in detail. Section \ref{sec:performance} presents the results of our comparison experiments and a discussion of the
influence of anode contact materials on the performance.  In section \ref{sec:Simulation}, we show that we can reproduce experimental data with computer simulations. For this, $\mu\tau$-products of electrons and of holes need to be known. We explain how we measured the $\mu\tau$-products of electrons and used simulations to derive the $\mu\tau$-products of holes. In section \ref{sec:T}, we present studies of the temperature dependence on the detector response and the electronic properties of CZT. 

\section{Substrate, detector fabrication and setup for performance measurements}
\label{sec:detectorfabrication}
As CZT substrate, we used a detector from Orbotech Medical 
Solutions Ltd. \cite{Orbotech} grown by the 
Modified High-Pressure Bridgman (MHB) process, 
delivered with a monolithic Indium (In) cathode, and  8$\times$8 Indium anodes (pixels).
The size of the wafer was 2.0$\times$2.0$\times$0.5~cm$^3$, the 
pitch of the pixels (anode contacts) 2.46~mm, and the gap between the pixels 0.6~mm. 
In order to minimize unknown parameters, we used a 
newly delivered CZT detector from which we
removed the original contacts and deposited new ones using the following procedure.

In a first step, the original contacts were removed by polishing with abrasive paper, 
followed by fine polishing with 0.5~$\mu$m 
particle size alumina suspension, and then by rinsing clean with pure methanol.
During the polishing process, the quality of the polished surface was 
constantly monitored with an optical microscope. Afterwards, 
the detector was etched for two minutes in a 1\% bromine-methanol solution and subsequently rinsed with methanol. 
Metalization was performed with an electron beam evaporator keeping the original geometry: pixel pitch 2.46~mm, gap between pixels 0.6~mm.
A temperature sensor was used to ensure that the substrate temperature 
never exceeded 100$^{\circ}$C.

\begin{figure}
\centering
\includegraphics[width=7cm]{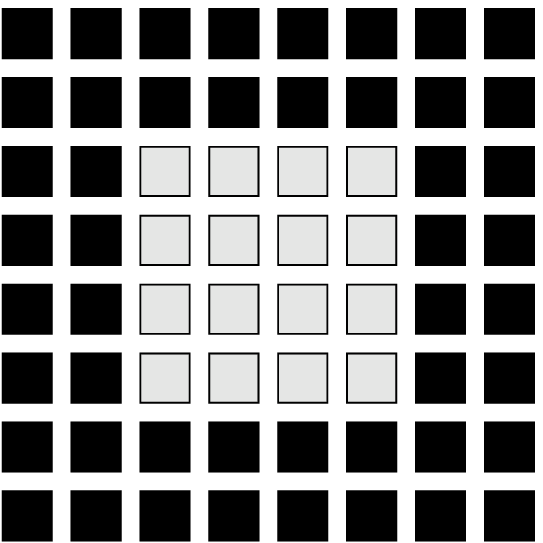}
\caption{Schematic view of the anode side of the CZT detector. For our study, we used the central 16 pixels, which are highlighted gray.}
\label{fig:pixelpattern}
\end{figure}

For the performance measurements, we used a custom-designed
PC card with 8$\times$8 contact pads, which had a Delrin \cite{Delrin} plastic fixture to mount the detector. Gold-plated, spring-loaded pogo-pins were used to 
connect the anode pixels to the PC Card, which fans out the contacts to
the readout electronics. The cathode was biased at -1000~V, while the anode 
pixels were held at ground potential. A hybrid electronic readout was used to read out the central 16 pixels
of the detector. 
Three of the pixels in the center of the crystal and the cathode (4~channels) 
were read out with a fast Amptek 250 amplifier. The post-amplified signals were digitized 
with a 4-channel 500 MHz oscilloscope, and transferred via Ethernet to a personal computer. 
This set-up allows us to measure the pulse length, and thus the drift times of the charge carriers
with an accuracy of 10 ns at 662~keV. The signals from the other 13 
pixels were read out with an ASIC\footnote{ev Products, ev-MultiPIX}  and a custom designed VME board. Therefore, we used only information of the central 16 pixels. Figure \ref{fig:pixelpattern} shows a schematic view of the anode side in which 
these pixels are highlighted gray. Additionally, we monitored the noise of the readout electronics with a pulser: both readout chains have a FWHM noise of $\sim$4~keV. Energy spectra were taken at 59~keV ($^{241}$Am), 122~keV ($^{57}$Co), and 662~keV ($^{137}$Cs).

\section{Substrate characterization}
\label{sec:char}

The Zn-fraction (x) in a Cd$_{1-x}$Zn${_x}$Te sample determines the bandgap and therefore the energy required for the production of electron-hole pairs. A 1\% variation of the zinc fraction corresponds roughly to a 0.5\% shift of the photopeak position. In the case of pixelated detectors, a variation in Zn fraction does not necessarily influence the performance, since each individual pixel's spectra can be corrected based on simple calibration measurements.  But, when we compare the performance of individual pixels on the same substrate, we have to exclude the possibility that local differences in crystal structure and composition are the cause of the observed effects. Therefore, the group at Fisk University used a photoluminescence method to map the zinc content, and infrared transmission microscopy to determine crystal defects of the CZT substrate. 

\subsection{Zn content photoluminescence mapping}
\label{subsec:Zn-map}
Photoluminescence (PL) mapping techniques \cite{Gfroerer:2000} normally use laser light having an energy above the bandgap of the crystal to illuminate the sample. Photons are absorbed and electronic excitations are created.  If through radiative relaxation light is emitted, the process is called photoluminescence.
For each spot on the crystal wafer, this emitted PL light is collected and fed into a spectrophotometer. The acquired PL spectra are fitted and the peak wavelength determined. From the peak position, the Zn-fraction (x) is calculated. \cite{Schlesinger:2001,Toney:96,Hjelt:97}.

Our PL mapping system operated at room temperature and used a 20~mW He-Ne laser ($\lambda$= 633~nm), a 10.16$\times$10.16~cm$^2$ X-Y stage, and a S2000 Miniature Fiber Optic Spectrometer from Ocean Optics. LabVIEW-based GUI software was developed by the group at Fisk University to perform automatic scanning, PL spectra acquisition, PL peak position analysis as well as to display in false color the Zn index (x) or Zn concentration on the surface of CZT sample. 

Sample preparation consisted of fine polishing without etching.

\begin{figure}
\centering
\includegraphics[width=10cm]{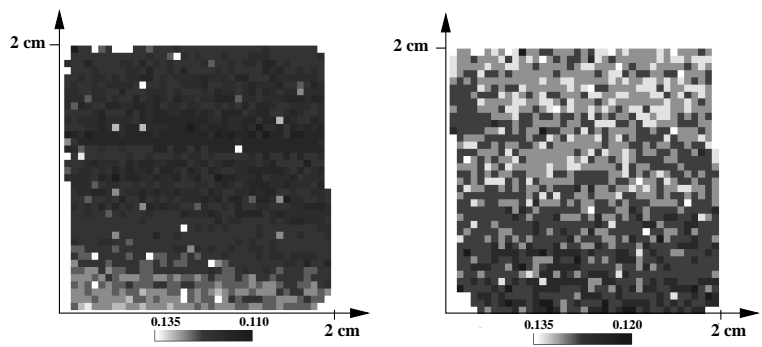}
\caption{The left panel shows the Zn fraction at the anode side coded on a gray scale from 0.110 to 0.135. On the right side the fraction at the cathode side is given with a gray scale from 0.120 to 0.135. }
\label{fig:znmapping}
\end{figure}

The results are shown in Fig. \ref{fig:znmapping}.
The Zn fraction  varies on the anode side between 0.110 and 0.135, on the cathode side between 0.120 and 0.135. 
Toney et al. \cite{Toney:96} performed similar room temperature PL mappings of HPB samples, which were  2~mm  thick and had an area of 1-2~cm$^2$.  These samples were etched in a 0.5\% bromine-methanol solution for 7~minutes. They showed a Zn fraction fluctuation of 1-2\% within one sample, and 5\%-10\% between the top and the heel of a 13~cm long boule. This variation is smaller than in our substrate. But, this may be due to the fact that our samples were not etched, since Li et al. \cite{Li:2001} reported that etching the substrates improves surface quality and reduces the Zn-fraction variations. 

\subsection{Detection of crystal defects with infrared transmission microscopy}
\label{subsec:defects}

\begin{figure}
\centering
\includegraphics[width=12cm]{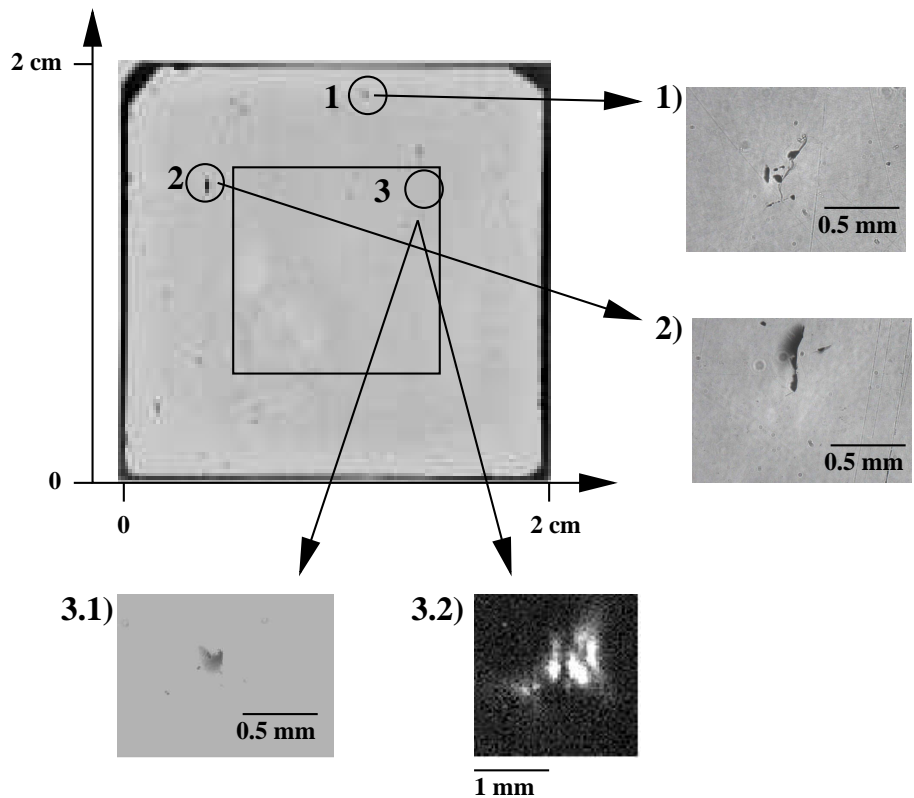}
\caption{Result of the infrared transmission microscopy measurement.
The upper left panel shows the entire substrate. The main defects are encircled, numbered, 
and presented in more detail on the righthand
panel. The square in the left
image marks the detector region we are using for our measurements.}
\label{fig:defects}
\end{figure}

Microscopy in infrared (IR) light of a wavelength, where CZT is sufficiently transparent, was used to localize crystal defects. In the case of CZT these are mostly Tellurium (Te) inclusions and grain boundaries. We used unpolarized IR light with a wavelength of 960~nm to reveal the general condition of the crystal. Polarized light with a wavelength of 1150~nm was used to determine the stressed regions in the CZT substrate. With a polarizer oriented at right angles with respect to the plane of polarization, unstressed material produces a dark field of vision. Because stress generates 
birefringence, stressed regions are seen bright on a dark background. In Fig. \ref{fig:defects} only Fig. 3.2 was taken with polarized light.

In Figure \ref{fig:defects}, left hand side, 
the image of the whole substrate is shown. There are some defects visible mainly in the outer region of the detector which we did not use for evaluation of detector performance. These defects are surface scratches caused by the detector holder and the gold plated pogo-pins (Fig. \ref{fig:defects}: 1, 2). In the central region relevant for our measurements, only one major defect is visible. This defect is shown with higher resolution in Fig. \ref{fig:defects}:  panels 3.1. (unpolarized) and 3.2 (polarized).  The defect is most likely a crack with Te inclusions. Using Fig. \ref{fig:defects}: 3.2 one finds that it covers a region of about 1 mm$^{2}$. It is localized about 200 micron below the anode side.
We will show later that the pixel in which the defect is located 
has an inferior performance compared to the others. We excluded this pixel from the analysis.

\section{Choice of contact materials and detector design for comparison tests}
\label{sec:contact}

As already mentioned in section \ref{Introduction}, one of the disadvantages of the MHB method is the lower 
bulk resistivity, compared to crystals grown with the HPB 
method, causing high leakage currents that contribute 
to the intrinsic noise and deteriorate
the performance of detectors at low energies ($<$~100~keV). 
The possibility to reduce bias currents by using high work-function (WF) blocking contacts on the cathode side of the n-type CZT was discussed by 
various authors (e.g.\ \cite{Nemirovski:01,Jung:05,Vadawale:04}). 

Nemirovski et al. \cite{Nemirovski:01} reported on the characterization of three types of contact configuration: first, ohmic In contacts on both, anode and cathode side, second, rectifying Au contacts on the cathode side with  
In contacts on the anode side, third, non-ohmic contacts on both, anode and cathode side (Au and Aluminum (Al)). At 122~keV, he obtained an energy resolution of 5.2\%, 4.6\%, and 7\% for the three configurations respectively. Vadawale et al. \cite{Vadawale:04} presented results of 20 
CZT substrates. The contacts were either standard In, Platinum (Pt), Au or hybrid contacts, i.e. Pt on the cathode side, and In, or Al on the anode side. Their best energy resolution was 5.0\% at 122~keV with Pt contacts followed by 5.7\% with Au contacts. They found also that the detector uniformity depends on the anode contact material: low work-function material on the anode side produces a more homogeneous 
response throughout the detector.

While these studies give interesting hints, it is still unknown which is the best choice of anode contacts. Therefore, we decided to study 
anode contact materials with a broad range of work-functions. We chose In, Ti, Cr, and Au with work-functions of 4.12~eV, 4.33~eV, 4.50~eV and 5.10~eV respectively for our comparison experiment.  
We excluded Pt (work-function: 6.35~eV) because substrates with Pt contacts showed poor performance when the deposition was done with the help of an electron beam system (for more details see \cite{Jung:05}).
As cathode contact material, we used Au, since our earlier study (see \cite{Jung:05}) has shown that Au reduces the dark current and improves the energy resolution.

\begin{figure}
\centering
\includegraphics[width=12cm]{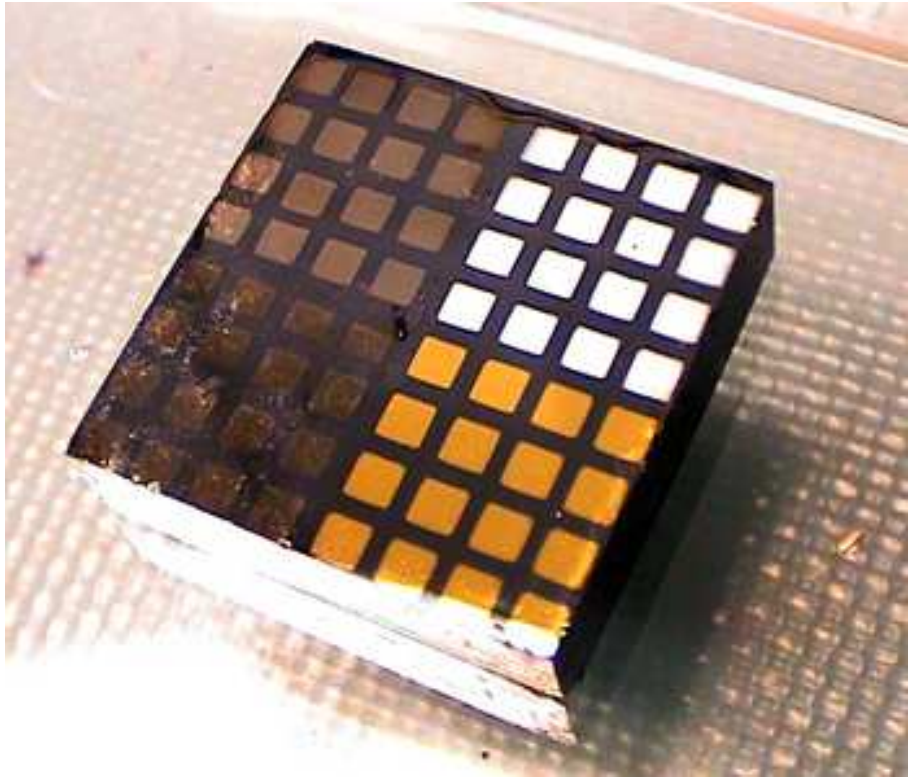}
\caption{The CZT detector (2.0$\times$2.0$\times$0.5~cm$^3$) with 8$\times$8 pixels (2.46~mm pitch)
and monolithic Au cathode. Starting clockwise from the top left quarter, the anode materials are Ti, In, Au, and Cr.}
\label{fig:contactmaterial}
\end{figure}

The surface and detector treatment before and during 
contact deposition influences the detector performance. Therefore we deposited the four different metals on a single detector substrate using a quarter of the 64 pixels for each material as shown in  Fig. \ref{fig:contactmaterial}. This design (4M-detector) ensures that all pixels have the same history. There remains another important point. One has to make sure that performance differences between different contact materials are not caused by substrate inhomogeneities. In section \ref{sec:char}, we described already some of our precautions. Additionally we fabricated a detector from the same substrate chip, where instead of four different materials only one, Ti, was used (Ti-detector). In this way we were able to compare pixels of the four sectors under identical conditions.

\section{Results of performance measurements comparing anode materials}
\label{sec:performance} 

The results discussed in this section relate to only 15 of the 16 central pixels. One pixel was removed from the analysis, because infrared microscopy (see Section \ref{sec:char}) has shown a major crystal defect in the area of this pixel leading to very poor performance in both contact configurations. Energy resolutions discussed in this section are resolutions of single pixels. Signals in neighboring pixels were ignored. For higher energies (122~keV and 662~keV) the anode signal was corrected for the depth of the photon interaction (DOI) based on the cathode-to-anode signal ratio.

Table \ref{tab:Am241} 
summarizes the results obtained with both contact configurations for an $^{241}$Am source. The  energy resolution is averaged over the three or four pixels of the 4M-detector with identical contact material and compared to an average over the corresponding pixels of the Ti-detector. The numbers are corrected for electronic noise. In parentheses, we added the 
uncorrected data. Both sets of data present the same picture. Therefore, we discuss the uncorrected data. Ti with resolutions of (4.2$\pm$0.2)~keV (4M-detector) as well as (4.4$\pm$0.2)~keV, (4.5$\pm$0.1)~keV, (4.3$\pm$0.2)~keV and (4.3$\pm$0.1)~keV (Ti-detector) had obviously the best performance followed closely by In (4.8$\pm$0.2)~keV. Au and Cr as anode contact materials are inferior by $\sim$8\%:  (5.3$\pm$0.2)~keV (Au), (5.4$\pm$0.3)~keV (Cr).

Table \ref{tab:MeanFWHM} summarizes  our results at 122 keV, and 662 keV. The important result is that all four different anode materials show rather similar energy dependence on the resolution, and that the order of performance remains unchanged.

\begin{figure}
\centering
\includegraphics[width=7cm]{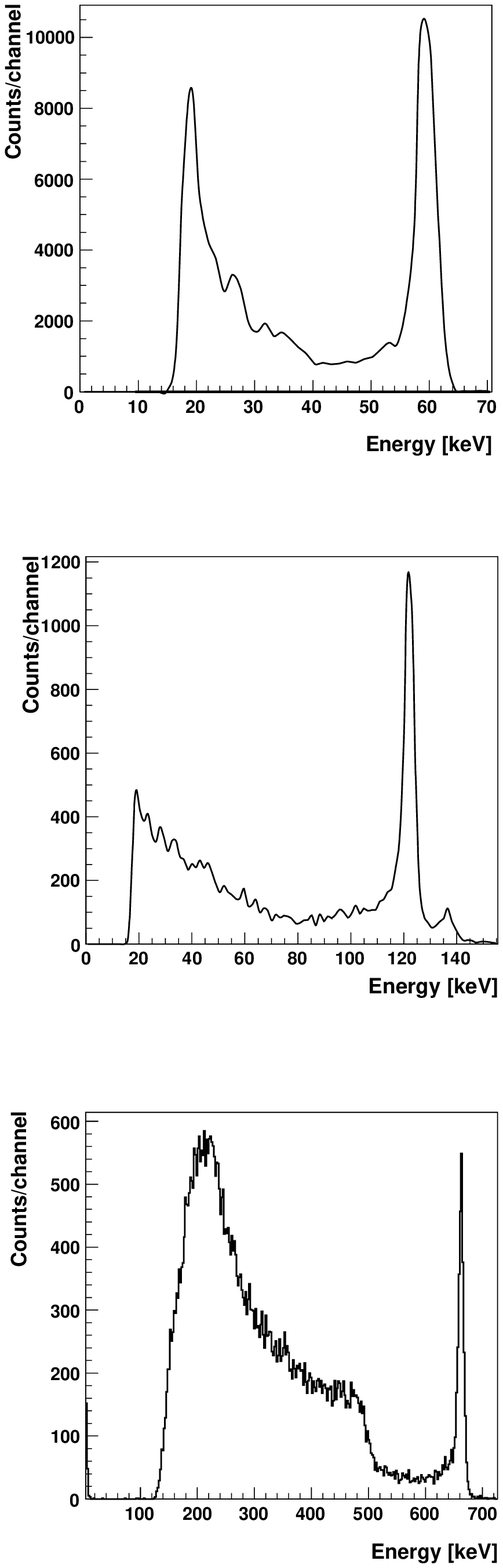}
\caption{From the top to the bottom, the panels show the spectra of a Ti pixel (Au cathode) 
at 59.5 keV ($^{241}$Am), 122.1 keV ($^{57}$Co), and 662 keV ($^{137}$Cs).}
\label{fig:spectra}
\end{figure}

In Fig. \ref{fig:spectra},  the energy spectra of a single Ti pixel are shown for three different photon energies. In case of the higher energies 122 keV ($^{57}$Co) and 662 keV ($^{137}$Cs),  the anode signal was corrected for the depth of the photon interaction (DOI) based on the cathode-to-anode signal ratio. Given the large pixel pitch of 2.46~mm, we obtained the excellent 
 energy resolution of 1.4~keV (4.0 keV), 2.0~keV (4.1 keV), and 7.2~keV (9.9 keV) for photon energies of 59~keV, 122~keV, and 662~keV, respectively. Again, the values given are corrected for electronic noise with the uncorrected data added in parentheses. 
The peak-to-valley ratio defined as the ratio of the maximal counts in the photopeak  to the mean of the channels between 560~keV and 580~keV, was $\sim$15.5 for 662~keV. 

Performance measurements for CZT crystals grown with the HPB process have been reported in the literature \cite{Zhang:05}. The energy resolution was 0.73\% and 0.93\% at 662~keV for two different detectors. Besides a different production process, there are other important features in which our setup varies from theirs. The substrate  measured 1.5$\times$1.5$\times$1.0~cm$^3$ with 11$\times$11 anode pixels and a pitch  1.27~mm. A steering grid between pixel anodes was used which was biased at negative voltage to focus the electrons to 
the anodes. The grid electrode was 100~$\mu$m wide and had a 200~$\mu$m gap between grid and  pixel.  These differences, one has to keep in mind when comparing both experiments, because these 
parameters have a strong influence on the energy resolution. A smaller pixel pitch and a larger thickness enhances the small pixel effect. Using simulations we have shown, that 
an increase in the thickness from 0.5~cm to 1~cm for a CZT detector with pixel pitch 1.64~mm will improve the energy resolution by $\sim$30\% (for more details see \cite{Jung:06}). Additionally 
we have compared simulated detectors with our geometry and the one used by Zhang et al. \cite{Zhang:05} but without a steering grid and found that the energy resolution is $\sim$40\%  better in the latter case. Taking this into account and the fact that a steering grid will further improve the resolution, we conclude that the resolution, we obtained, is comparable to the energy resolutions reported by Zhang et al.\cite{Zhang:05} of 5.0~keV and 6.2~keV at 662~keV.

\section{Detector simulations}
\label{sec:Simulation}
In this section, we first describe 3-D simulations of the detector response.
Subsequently, we present measurements and detailed comparisons of the simulated and
experimentally measured detector response which allow us to determine 
the $\mu\tau$-products for electrons and holes. 

We simulated a 2.0$\times$2.0$\times$0.5~cm$^3$ detector with 8$\times$8 pixels
with a pixel pitch of 2.46~mm  and 
a monolithic cathode. The simulation package consists of three 
components (for further details see \cite{Jung:06}). The first component 
calculates with a finite-difference method the 3-D electric field as well as the weighting potentials of the 
pixel and cathode contacts. In the second, 
the Geant 4.0 code \cite{G4} is used to simulate the interactions of the incident 
$\gamma-$rays and the secondaries produced in the detector.
The third component uses the information, where the primary photons deposit  
energy in the detector to generate electrons and holes, and tracks them through the detector.
This component uses the electric field and the weighting potentials to compute the charge induced on the contacts. 
Electrons reaching the substrate surface in the gap between the pixels were left at the position, where they hit the surface, i.e. the surface conductivity was set to infinity. 
The detector material was assumed to be uniform. We measured electron mobility as well as  electron  lifetime and used these values as input parameters in our simulations. Hole mobilities and lifetimes  were determined from comparisons of simulated and experimental data.

\begin{figure}
\centering
\includegraphics[width=14cm]{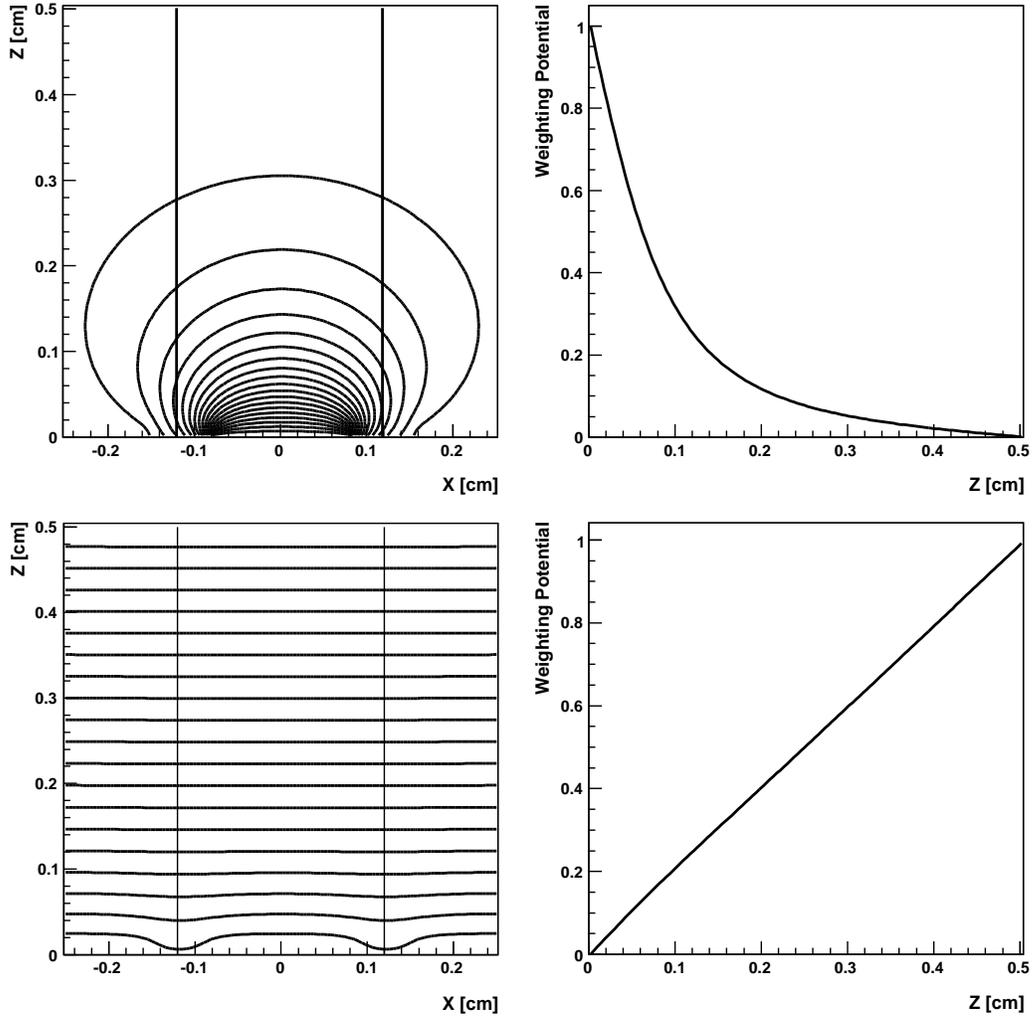}
\caption{Results of the detector simulations. The two panels on the left side show cross 
sections of the weighting potential of an anode pixel (top) and the cathode contact (bottom). 
The anode pixels are located at the bottom sides of the two panels, and the vertical lines 
indicate the location and width of the central pixel. The two panels on the right side, 
show how the weighting potentials depend on the height $z$ above the anode pixel.
In most applications the orientation of the detectors lets radiation enter from the
cathode side.}
\label{fig:WP}
\end{figure}

The weighting potentials of a central pixel and the cathode contact are presented in Figure \ref{fig:WP}.
The graphs show a small-pixel-effect \cite{Barr:95,Luke:95}, i.e. the pixel weighting potential
has a steep gradient close to the pixel. As the pixel pitch (2.46~mm) is roughly comparable
to the detector thickness (5~mm), the small-pixel-effect is not very pronounced. Therefore,  electrons and holes induce charge on the anode, even if they move deep inside the detector.
The cathode signal does not show any small-pixel-effect and depends almost linearly 
on the distance from the cathode contact. 

The first step in the determination of electron mobility and lifetime is the measurement of the electron $\mu\tau$-product ($\mu_e\, \tau_e$), which can be calculated 
applying formula \ref{mutau}:
\begin{equation} 
  \label{mutau}
  \mu_e \tau_e = \frac{d^2}{\ln{\frac{M_{b1}}{M_{b2}}}}(\frac{1}{V_2}-\frac{1}{V_1})
\end{equation}
where $d$ is the detector thickness, and $M_{b1}$ and $M_{b2}$ are the photopeak centroids 
for two different bias voltages $V_1$ and  $V_2$ \cite{He:98}. A low energy source has to be used to ensure that most of the interaction occurs near the cathode, and that hole contributions can be neglected.

\begin{figure}
\centering
\includegraphics[width=14cm]{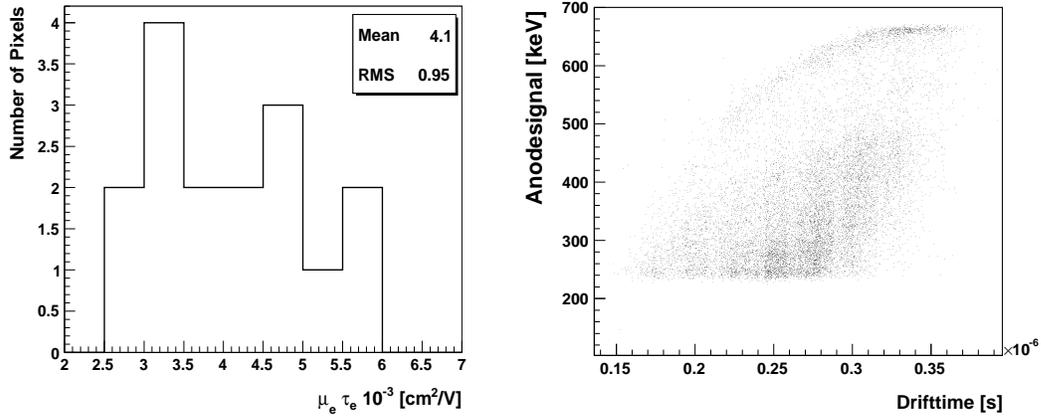}
\caption{The left panel shows the measured $\mu_e \tau_e$-product for the 16 
central pixels. The right panel displays the anode charge as function of pulse 
width (duration) for a cathode bias of -1000~V. The pulse width is largely 
given by the drift time of the electrons 
through the detector.}
\label{fig:SimParameter}
\end{figure}

The left panel of Fig. \ref{fig:SimParameter} 
shows the $\mu_e\,\tau_e$-product for the central 16 pixels. 
The mean is $4.1 \cdot 10^{-3}$~cm$^2$ V$^{-1}$ and the RMS is 0.95. \\
The $\mu_e\,\tau_e$-product of HPB CZT substrates are given
by F. Zhang et al. \cite{Zhang:05}  and by S. Barthelmy et al. \cite{swift}.
F. Zhang et al. \cite{Zhang:05} reported the $\mu_e\,\tau_e$-product of two different HPB CZT substrates with means of $3.22 \cdot 10^{-3}$~cm$^2$ V$^{-1}$ and of $6.79 \cdot 10^{-3}$~cm$^2$ $V^{-1}$ and RMS values of $0.22 \cdot 10^{-3}$~cm$^2$ $V^{-1}$ and $0.51 \cdot 10^{-3}$~cm$^2$ $V^{-1}$ respectively, which shows that CZT detectors grown with the MHB process have a similar mean, but a wider spread. \\
The $ \mu_e\,\tau_e$-products of all 32,768 planar CZT HPB detectors (4$\times$4$\times$2 mm$^3$) used in the Burst Alert Telescope(BAT) on-board the Swift Gamma-ray Burst Explorer (see section \ref{Introduction}) were measured \cite{swift}, and they vary 
between 5.0$\times$10$^{-4}$ and 1.0$\times$10$^{-2}$ cm$^2$ V$^{-1}$. The variation within a single detector was not determined. These measurements show that the two detectors used by Zhang et al. have a $\mu_e\,\tau_e$-product above the average value of HPB CZT detectors. 

The electron mobilities can be inferred from the widths (durations) of the induced anode 
pulses, which are mainly determined by the drift times of the electrons.
The right side of Fig.~\ref{fig:SimParameter} shows the correlation of the pulse widths
and the induced anode charge. With a detector bias of -1000 V, the longest observed drift times
were $\simeq$0.35~$\mu$sec. Thus, given that electrons drift up to 0.5~cm through the
detector, we determined an electron mobility of $\mu_e = 695$~cm$^2$ V$^{-1}$ s$^{-1}$ and a mean electron lifetime of $\tau_e = 5.9 \cdot 10^{-6}$~s using the previously measured mean $\mu_e\,\tau_e$-product. 

\begin{figure}
\centering
\includegraphics[width=14cm]{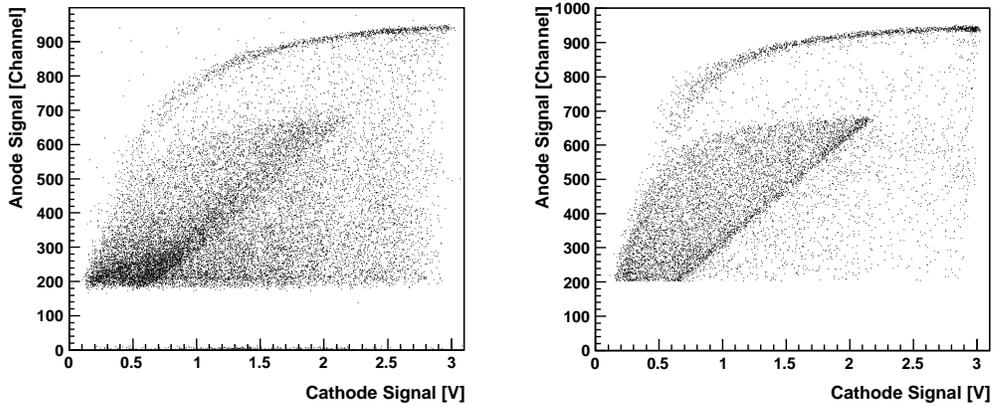}
\caption{The anode-cathode charge correlation for experimental (left side) and
simulated (right side) data. The simulations used the following input 
parameters: 
$\mu_e = 695$~cm$^2$ V$^{-1}$ s$^{-1}$, $\tau_e = 5.9 \cdot 10^{-6}$~s, 
$\mu_h = 50$~cm$^2$ V$^{-1}$ s$^{-1}$, and $\tau_h = 1 \cdot 10^{-6}$~s.}
\label{fig:VergleichSim_Data}
\end{figure}

With these electron $\mu_e$ and $\tau_e$-values, we deducted the corresponding values for holes
by comparing the measured anode to cathode correlation with simulated correlations for different
$\mu_h$ and $\tau_h$. Fig. \ref{fig:VergleichSim_Data} right side displays the simulated correlation with $\mu_h = 50$~cm$^2$ V$^{-1}$ s$^{-1}$ and $\tau_h = 1 \cdot 10^{-6}$~s and for comparison the corresponding correlation of our experimental data (left side). 
Even the finer details of the experimental data plot are reproduced well.
We used this set of parameters in all simulations. The following paragraphs will show that simulations based on these parameters reproduce the experimental data.

\begin{figure}
\centering
\includegraphics[width=14cm]{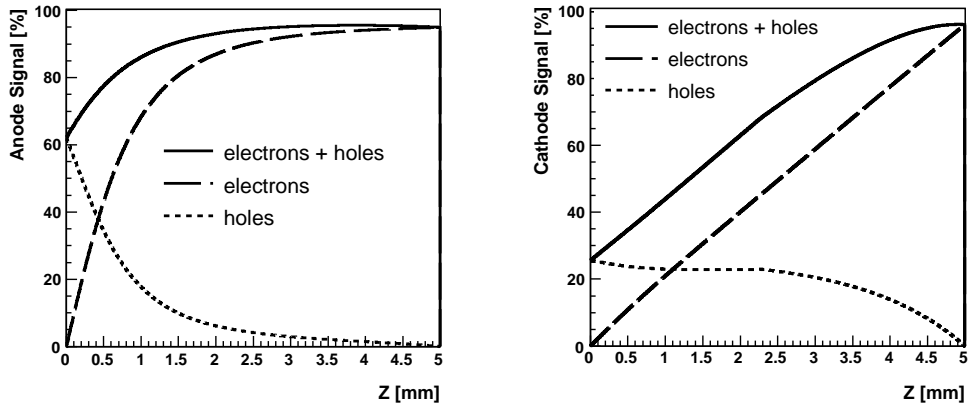}
\caption{The left (right) panel shows the simulated anode (cathode) 
signal as a function of the distance of energy deposition from the anode  (Z-direction). 
The solid lines gives the total signal, the long-dashed line the electron contribution, 
and the dotted line the hole contribution. Z is the distance from the anode. 100\% anode (cathode) signal means, that the whole deposited energy is measured.}
\label{fig:CathodeSignal}
\end{figure}

Having determined the hole parameters, we can study separately the contributions of electrons and holes to the anode and cathode signals (see Fig. \ref{fig:CathodeSignal}). The anode signal dependency on the distance of energy deposition from the anode is given in the left panel of Fig. \ref{fig:CathodeSignal}. The signal drops close to the anode due to the steep weighting potential increase. In this area the hole contribution is maximal and decreases to almost zero close to the cathode. The cathode signal dependency is given on the right panel. It is almost linear with the 
distance from the anode side. The deviation close to the cathode is caused by holes reaching the cathode before being trapped. 
The approximately linear dependence of the induced cathode charge on the 
depth of the photon interaction (DOI) can be used as a diagnostic tool 
to test, whether the substrate is uniform.

We compared the behavior of the signal dependency on the DOI by dividing  
the total cathode signal into 
five different signal ranges (0.8-1.2~V, 1.2-1.6~V, 1.6-2.0~V, 
2.0-2.4~V and 2.4-2.8~V). For each interval, the position of the photopeak 
and the number of events in the photopeak was determined separately.
The number of photopeak events is defined as the number of events lying between $\pm\:3 \cdot \sigma$ around the photopeak position. 

\begin{figure}
\centering
\includegraphics[width=6in]{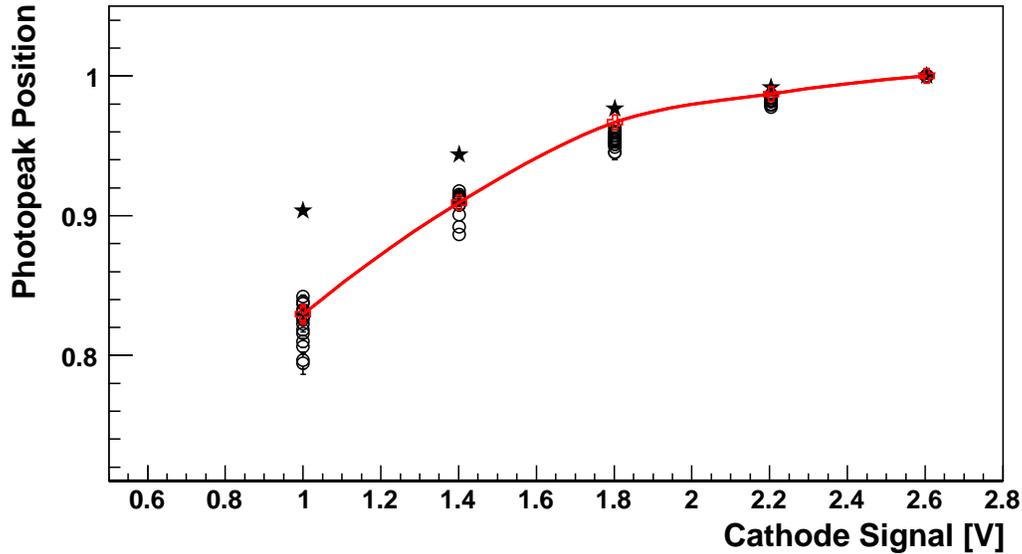}
\caption{The mean photopeak position for five different cathode signal 
ranges (0.8-1.2~V, 1.2-1.6~V, 1.6-2.0~V, 2.0-2.4~V and 2.4-2.8~V) and for four
different anode materials. For each material four pixels were measured 
separately. The results for simulated data are shown as open crosses 
connected with a line.  The photopeak position was normalized to one for the fifth 
range and all other ranges scaled accordingly. All pixels show the same 
behavior with the exception of one titanium pixel (filled stars).}
\label{fig:DOIDependenc1}
\end{figure}

The correlation of the anode versus the cathode signals depends on 
where in the crystal the charge is generated and to where it drifts. 
Figure \ref{fig:DOIDependenc1} shows the measured and simulated
mean photopeak positions for different intervals of the cathode signal. 
While the measured data consists of four different sets for each of the four materials, there is only one set of simulated data, because the contact material was not taken into account in our simulations. 
The agreement between data and simulations is excellent for 
all but one pixel. This pixel is the one which was excluded from the 
analysis, because 
infrared transmission microscopy showed the presence of a 
crystal defect close to the anode (section\ \ref{subsec:defects}).

\begin{figure}
\centering
\includegraphics[width=6in]{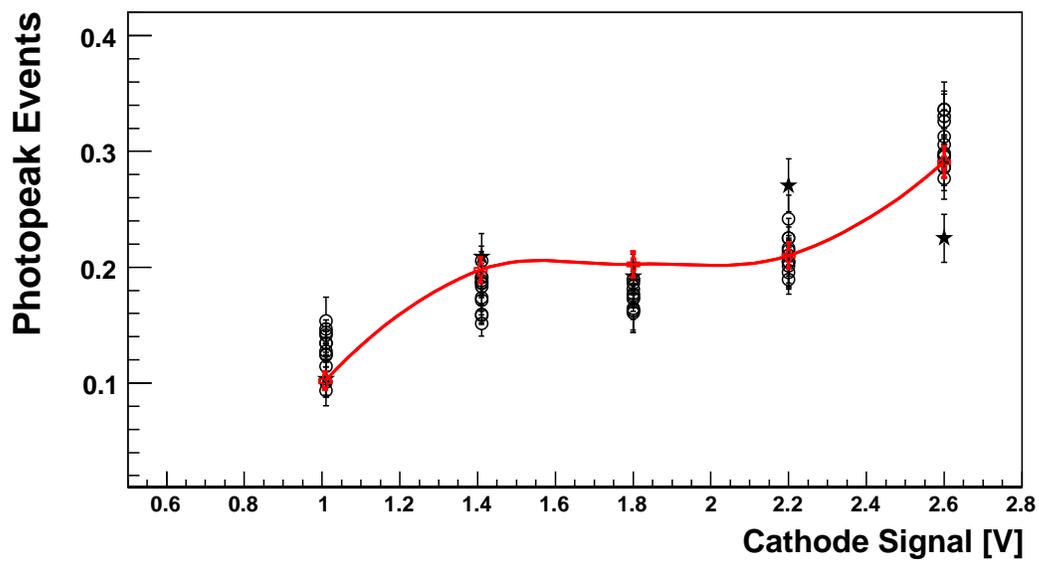}
\caption{The number of events in the  photopeak for five different 
cathode signal ranges (0.8-1.2~V, 1.2-1.6~V, 1.6-2.0~V, 2.0-2.4~V and 
2.4-2.8~V) 
and for four different anode materials. For each material the pixels were 
measured separately. The simulated data are given as open crosses and 
connected with a line.
The number of events is normalized to the total number of 
photopeak events in all ranges for each metal. }
\label{fig:DOIDependenc2}
\end{figure}

Figure \ref{fig:DOIDependenc2} compares the measured and the simulated relative numbers of photopeak events for different ranges of the cathode signal.
The measured photopeak events distribute as in the simulations, one pixel deviating more than the others. For this pixel, 
infrared transmission microscopy shows the presence of a crystal defect 
close to the anode (see Sect. \ref{subsec:defects}).

Because the simulation matches the experimental data well, 
we conclude that in both cases for a given cathode signal interval the 
charge generation occurs in the same range of DOIs for simulation and experiment. For the deviating pixel (the
same as above), the mean photopeak position varies less than for the others,
confirming that the crystal defect affects the charge induced at the cathode
and/or the anode.
We conclude that the full volume under all but one pixel 
is active detector volume and produces proper cathode signals.


\section{Temperature dependence}
\label{sec:T}

\begin{figure}
\centering
\includegraphics[width=10cm]{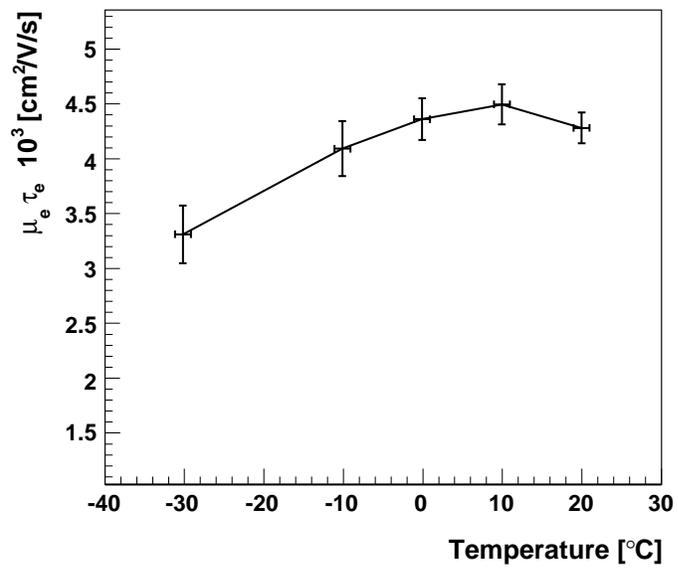}
\caption{Mean temperature dependence of the $\mu_e \, \tau_e$-products 
for the central 16 pixels. The errors given are the errors on the mean 
value.}
\label{fig:TempAbhaengigkeit}
\end{figure}

In satellite experiments ambient air temperature is well below $0^{\circ}$C. Therefore we studied the dependence of the electronic properties and the performance of
the CZT detector over a broad temperature range from room temperature down 
to -30$^{\circ}$C. One important parameter is the $ \mu_e\,\tau_e$-product. It influences the detector performance. Decreasing product means inferior resolution.
Therefore we measured the electron $\mu\tau$-product and the spectra as function of temperature 
with an $^{241}$Am source as described in section \ref{sec:Simulation}.
Figure \ref{fig:TempAbhaengigkeit} gives the results. The product shows the
highest values at and above 0$^{\circ}$C and decreases at lower temperatures by up to 
$\sim$30\%. Similar behavior has been reported for HPB substrates \cite{Stur:05} which had a maximum at 5$^{\circ}$C with $\mu_e\, \tau_e$$\sim$~9$\times$10$^{-3}$cm$^{2}$ V$^{-1}$ with a drop of about $\sim$30\% at -20$^{\circ}$ to $\mu_e\, \tau_e$$\sim$~6$\times$10$^{-3}$cm$^2$ V$^{-1}$. 

\begin{figure}
\centering
\includegraphics[width=15cm]{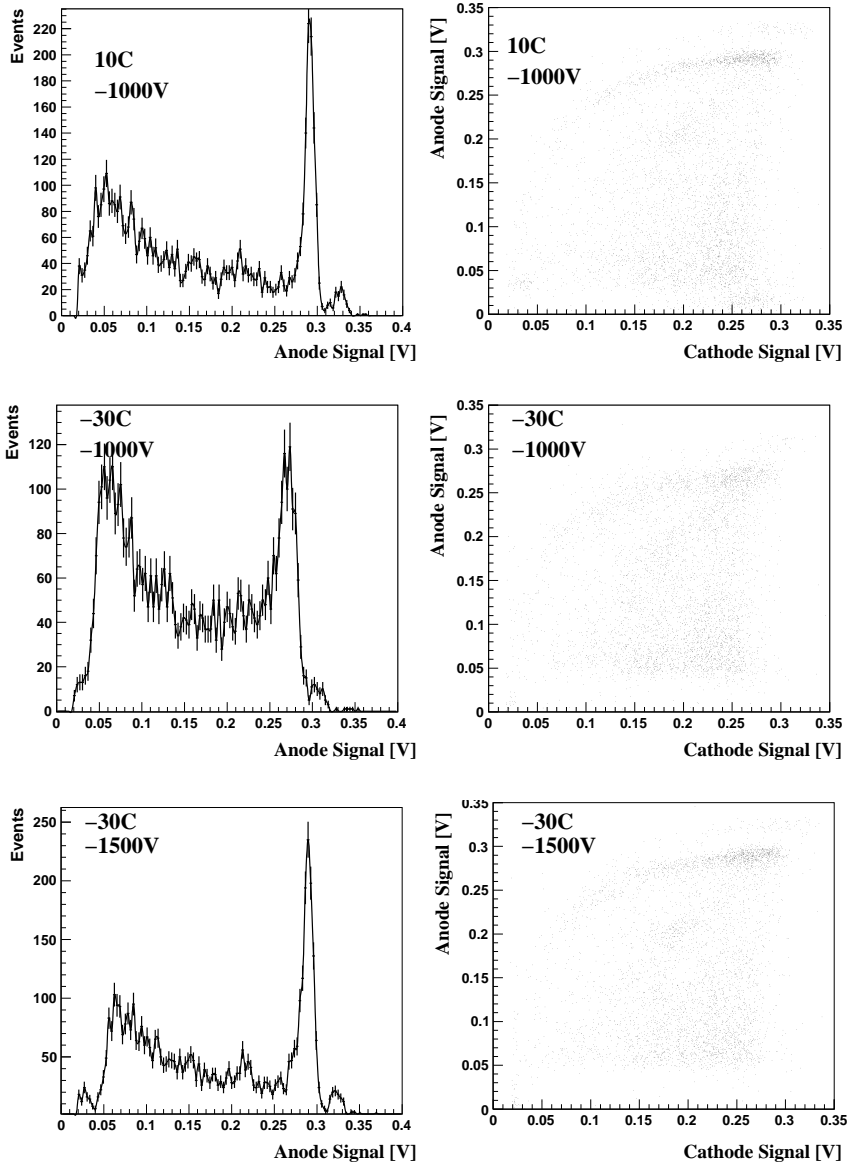}
\caption{Energy spectra at 122~keV ($^{57}$Co source) and the 
anode signal dependence on cathode signal with a cathode bias voltage of 
-1000~V for two different temperatures
(10$^{\circ}$C, top, and -30$^{\circ}$C, middle). The signals were measured with the Amptek 250 readout chain and are therefore given as voltages. At the bottom the spectrum 
with a bias of -1500~V at a temperature of -30$^{\circ}$C, as in the center
plot. The energy resolutions given  are 4.4\%, 10.6\% and 5.5\% FWHM (top to 
bottom).}
\label{fig:temp_spectra}
\end{figure}

In Fig.~\ref{fig:temp_spectra} 122 keV 
energy spectra are presented. They are measured at 10$^{\circ}$C and at -30$^{\circ}$C together with their anode to cathode signal correlations. The top two panels show that the energy resolution 
deteriorates markedly, when the detector is operated at lower temperatures and all other conditions (bias voltage) are kept constant. The anode-cathode correlation shows that the width of 
the photopeak increases for all cathode signal amplitudes, and thus for all 
DOIs. 
In the measurements that just have been discussed, we kept the bias voltage constant at 
-1000~V. This voltage has been chosen, because we found that at temperatures above 
10$^{\circ}$C a change of voltage from -1000~V to  -1500~V will increase the 
dark current by several orders of magnitude,  so much, that no useful spectra could be obtained.
 If the detector is cooled, 
however, this electrical 
breakdown phenomenon is suppressed, and higher bias voltages become possible. 
The bottom panel of Fig.~\ref{fig:temp_spectra} shows an 
energy spectrum measured at -30$^{\circ}$C and a cathode bias of -1500~V.
With the increased bias voltage most of the room temperature performance 
is recovered: energy resolution of 5.5\% at -30$^{\circ}$C compared to 4.4\% at 10$^{\circ}$C,   and a somewhat lower signal amplitude (0.28 V and to 0.29~V respectively). 

While most of the performance degradation at lower temperatures will be due to the reduction of the $ \mu_e\,\tau_e$-product, another effect may have influenced our measurements to some extent:  vibrations of the temperature chamber which was used to cool the
CZT detector.

\section{Conclusion}
\label{sec:disc}
In this paper, we present the results of a detailed study of the performance 
of a MHB CZT detector from the company Orbotech Medical Solutions Ltd. \cite{Orbotech}. \\
Using photoluminescence, we mapped the spatial distribution of the zinc content, 
which showed a variation of  $\sim$3\%. This variation does not affect the
energy resolution of the detector if information about the location of
the interaction is available (e.g. pixelated detectors), because 
in this case the effect can be corrected for.
With infrared transmission microscopy, we could detect some crystal defects and 
could correlate poor pixel performance with such defects. \\
We tested various anode contact materials and found that Ti showed the best performance
of 2.0~keV, 1.9~keV and 7.3~keV at 59~keV, 122~keV and 662~keV, respectively.
We are using this metal now 
to contact all our MHB detectors.
We showed detailed comparisons of experimentally measured and simulated 
detector response. After adjusting the hole mobility and lifetime, our simulation gives 
a good description of the measured data and shows that we have a deep understanding
of our data. Therefore, the simulation can be used to optimize the 
detector design as  the pixel width and pitch.\\
We studied the detector properties as a function of temperature. The electron 
$\mu\,\tau$-product drops by about 30\%, when cooling the substrate from 
20$^{\circ}$C to -30$^{\circ}$C with constant bias voltage, and the energy resolution deteriorates 
accordingly, a behavior that has been noted already for MHB substrates \cite{Nari:00}.
For HPB substrates, a corresponding decrease of the electron $\mu\,\tau$-product 
has been observed \cite{Stur:05}.
We were able to recover almost all of the room temperature performance by increasing 
the bias voltage at low temperatures, which is possible, because at lower temperatures the breakdown voltage is increased. 
Taking everything into account, we conclude, that the overall performance of CZT detectors, produced with the cost-effective MHB process, is comparable to the performance of CZT detectors which are produced with the expensive HPB process.
\\[2ex]
{\it Acknowledgments:}
We thank Uri El Hanany from Orbotech Inc. for several free CZT detectors. 
We acknowledge S. Komarov, L.~Sobotka, D.~Leopold, and J.~Buckley for helpful discussions. 
Thanks to electrical engineer P.~Dowkontt, and electrical technician 
G.~Simburger for their support.
This work is supported by NASA under contracts NNG04WC176 and 
NNG04GD70G, and the NSF/HRD grant no.\ 0420516 (CREST).
The authors at Fisk University gratefully acknowledge financial support
from the National Science Foundation through the Fisk University Center
for Physics and Chemistry of Materials (CPCoM), Cooperative Agreement
CA: HRD-0420516 (CREST program), and from US DOE through the National
Nuclear Security Administration (NNSA), Office of Nonproliferation
Research and Engineering (NA-22), grant no. DE-FG52-05NA27035.

\newpage

\clearpage
\newpage

\begin{table}[tb]
  \centering
  \begin{tabular}[h]{ l l l l}
    \hline
    \multicolumn{4}{ l}{Energy Resolution (keV)}\\  \hline
    \multicolumn{4}{ l}{59.5 keV } \\  \hline
   \multicolumn{2}{l}{Ti-detector} & \multicolumn{2}{l}{4M-detector}\\ 
   Ti & 2.1$\pm$0.2 (4.4$\pm$0.2) & Au &  3.5$\pm$0.3 (5.3$\pm$0.2)  \\ 
   Ti & 2.1$\pm$0.2 (4.5$\pm$0.1) & Cr &  3.1$\pm$0.1 (5.4$\pm$0.3)   \\ 
   Ti & 2.0.$\pm$0.2 (4.3$\pm$0.2)  & In &  2.1$\pm$0.2 (4.8$\pm$0.2) \\ 
   Ti & 2.0$\pm$0.1 (4.3$\pm$0.1)  & Ti &  1.9$\pm$0.2 (4.2$\pm$0.2) \\ 
 \end{tabular}
  \caption{For both configurations (Ti­detector and 4M­detector) and for each anode material, the mean value and the error are given. In parentheses, values which have not been corrected for electronic noise. Each row corresponds to the same area of of the CZT substrate. The values in the first three rows are determined with four pixels. In the fourth, only three pixels were used, since one was excluded because of crystal defects (see Section 3.2). 
} 
\label{tab:Am241}
\end{table}

\begin{table}[tb]
  \centering
  \begin{tabular}[h]{l l l}\hline
    \multicolumn{3}{l}{Energy Resolution (keV)}\\ \hline
     & 122.1 keV & 662 keV \\ \hline 
    Au & 3.3$\pm$0.3 (7.1$\pm$0.2) & 9.2$\pm$0.1 (11.8$\pm$0.2) \\ 
    Cr & 4.3$\pm$0.1 (6.5$\pm$0.4)& 8.6$\pm$0.6 (11.6$\pm$0.4) \\ 
    In & 2.9$\pm$0.1 (5.4$\pm$0.1) & 9.4$\pm$0.2 (12.6$\pm$0.3) \\ 
    Ti & 1.9$\pm$0.1 (4.7$\pm$0.3) & 7.4$\pm$0.3 (10.4$\pm$0.3) \\ \hline
 \end{tabular}
  \caption{Energy resolutions (FWHM) obtained with pixels made of different metals. For each anode 
material the mean value and the error on the mean value are given. In parentheses, 
values which have not been corrected for electronic noise. The values in the first three lines are determined with four pixels. In the fourth, only three pixels were used, since one was excluded because of crystal defects (see Section \ref{subsec:defects}).} 
\label{tab:MeanFWHM}
\end{table}


\begin{thebibliography}{00}
\bibitem{Orbotech}{Orbotech Medical Solutions Ltd., 10 Plaut St., Park Rabin, P.O.Box: 2489, Rehovot, Israel, 76124}
\bibitem{swift}{S. D. Barthelmy, Proc. SPIE, 5165 (2004) 175 }
\bibitem{Grindlay}{J. E. Grindlay  \& the EXIST Team 2005, New Astron. Rev., 49 (2005) 435 }
\bibitem{Nemirovski:01}{Y. Nemirovski, G. Asa, A. Peyser, NIM, A 458 (2001) 325-333}
\bibitem{Vadawale:04}{S. V. Vadawale et al., Proc. SPIE, 5540 (2004) 22-32 }
\bibitem{Nari:98}{T. Narita, P. Bloser, J. Grindlay et al., Proc. SPIE, 3446 (1998) 218 }
\bibitem{Nari:00}{T. Narita, P. Bloser, J. Grindlay, J.A. Jenkins, Proc. SPIE, 4141 (2000) 89}
\bibitem{Nari:02}{T. Narita, J.E. Grindlay, J.A. Jenkins et al.,  Proc. SPIE, 4497 (2002) 79}
\bibitem{Jung:05}{I. Jung, M. Groza, J. Perkins, H. Krawczynski, A. Burger,  Proc. SPIE, 5922, (2005) }
\bibitem{Delrin}{http://heritage.dupont.com/floater/fl\_delrin/floater.shtml}
\bibitem{Gfroerer:2000}{T. H. Gfroerer, in: R.A. Meyers (Ed.), Encyclopedia of Analytical Chemistry, John Wiley, Chichester, (2000), pp. 9209-9231}
\bibitem{Schlesinger:2001}{T. E. Schlesinger et al., Mat. Sci. Eng., R 32 (2001) 103-189} 
\bibitem{Toney:96}{J. E. Toney et al., NIM, A 380 (1996) 132}
\bibitem{Hjelt:97}{K. Hjelt et al., Phys. Stat. Sol., (a) 162 (1997) 747 }
\bibitem{Li:2001}{Z.-F. Li et al.,  J. Appl. Phys., 90 (2001) 260-264  }
\bibitem{Zhang:05}{F. Zhang et al., IEEE Trans. Nucl. Sci., 52 (2005) 2009-2016}\bibitem{Jung:06}{I. Jung, H. Krawczynski, S. Komarov,L. Sobotka, Astrop. Phys., 26 (2006) 119}
\bibitem{G4}{S. Agostinelli et al., NIM, A 506 (2003) 250-303 }
\bibitem{Barr:95}{H.~H.Barret, J.~D. Eskin, H.~B. Barber, Phys. Rev. Lett., 75 (1995) 156}
\bibitem{Luke:95}{P. N. Luke, in: Proc. of the ``9th International Workshop
on Room-Temperature Semiconductor X- and Gamma-Ray Detectors, Associated Electronics and
Applications'', Grenoble, (1995)}
\bibitem{He:98}{Z. He, G. F. Knoll, D. K. Wehe, J. Appl. Phys., 84 (1998) 5566}

\bibitem{Stur:05}{B. W. Sturm, Z. He, T. H. Zurbuchen, P. L. Koehn, IEEE Trans. Nucl. Sci., 52 (2005) 2068-2075}

\end{thebibliography}
\end{document}